\documentstyle[a4,12pt,amssymb,epsfig]{article}
\begin{document}
\textheight 23cm 
\topmargin -2.4cm
\textwidth 16cm
\oddsidemargin 0.65cm 
\evensidemargin 0.65cm
\setlength{\baselineskip}{.75cm}
\setlength{\parskip}{0.45cm}
\begin{titlepage}
\begin{flushright}
\begin{tabular}{l}
CERN-TH/97-280 \\
hep-ph/9712273 
\end{tabular}
\end{flushright}
\vspace*{1cm}
\begin{center}
\Large{
{\bf Next-to-leading order QCD corrections to\\} 
\vspace*{0.5cm}
{\bf inclusive-hadron photoproduction in\\}
\vspace*{0.5cm}
{\bf polarized lepton-proton collisions\\}}
\vskip 2.5cm
{\large D.\ de Florian and W.\ Vogelsang}  \\
\vspace*{0.7cm}
\normalsize
Theory Division, CERN,\\
CH-1211 Geneva 23, Switzerland \\
\end{center}
\vskip 1.2cm
\begin{center}
{\bf Abstract} \\
\end{center}
\vspace{0.5cm}
\normalsize
We calculate the next-to-leading order QCD corrections to the ``direct'' part
of the spin-dependent cross section for single-inclusive charged-hadron 
photoproduction. This process could be studied experimentally in future 
polarized fixed-target lepton-nucleon experiments, but also at the HERA 
$ep$ collider after an upgrade to both beams being polarized. 
We present a brief numerical evaluation of our results by studying
the $K$-factors and the scale dependence of the NLO cross section. 
CERN-TH/97-280 \\
December 1997 
\end{titlepage}
\section{Introduction}
In the last few years measurements of the spin asymmetries 
$A_1^N$ ($N=p,n,d$) in longitudinally polarized deep-inelastic scattering 
(DIS) have provided much new information on the spin structure of the 
nucleon. Theoretical leading order (LO) [1--3] and
next-to-leading order (NLO) [1--4] analyses of the data
sets demonstrate, however, that these are not sufficient to accurately
extract the spin-dependent quark ($\Delta q = q^{\uparrow}-q^{\downarrow}$) 
and gluon ($\Delta g=g^{\uparrow}-g^{\downarrow}$) densities of the nucleon. 
This is true in particular for $\Delta g(x,Q^2)$ since it contributes to 
DIS in LO only via the $Q^2$-dependence of $g_1$ (or $A_1$) 
which could not yet be 
accurately studied experimentally. As a result of this, it turns out 
[1--4] that the $x$-shape of $\Delta g$ seems to be hardly 
constrained at all by the DIS data, even though a tendency towards a 
fairly large positive {\em total} gluon polarization, $\int_0^1 \Delta 
g(x,Q^2=4 \; \mbox{GeV}^2) dx \gtrsim 1$, was found \cite{grsv,gs,bfr}. 
The measurement of $\Delta g$ thus remains one of the most interesting 
challenges for future spin physics experiments. 
When selecting suitable processes for a determination of $\Delta g$,
it is crucial to pick those that, unlike $g_1$, have a gluonic
contribution already at the lowest order. Sticking to polarized lepton-nucleon
interactions, this implies to consider processes less inclusive than DIS.
Among those is the production of a (charged) hadron with large transverse 
momentum $p_T$. To obtain a large number of such hadrons, it is 
expedient to go to {\em photo}production, i.e. to the limit when the
(circularly polarized) photon which is exchanged between the polarized lepton 
and the nucleon, is almost on-shell. In this way one avoids the suppression 
of the cross section by the photon propagator. 

As was shown recently \cite{sv}, a 
polarized version of the HERA collider with $\sqrt{s}\approx 300$ GeV 
would be a very promising and useful facility for studying polarized 
photoproduction reactions. In particular, two of the conceivable processes, 
single-inclusive hadron production and jet production, show strong sensitivity
to the polarized gluon distribution of the proton and also appear 
likely to yield statistics good enough for a successful measurement~\cite{sv}.
In the framework of the LO calculation performed in \cite{sv}, the 
sensitivity of 
these reactions to $\Delta g$ is due in the first place to the subprocess 
$\vec{\gamma} \vec{g} \rightarrow q\bar{q}$, where the arrows denote 
longitudinal polarization. As was stressed in \cite{sv}, and as is 
well-established in the unpolarized case, the (quasi-real) photon will not only
interact in a direct (``point-like'') way, but can also be resolved into 
its hadronic structure. As far as a determination of $\Delta g$ is
concerned, such ``resolved'' contributions (which appear at the same order 
in perturbation theory as the ``direct'' piece) are to be considered as a 
background. As was shown in \cite{sv}, the resolved component is 
subdominant with respect to the direct one in certain regions of 
rapidity and transverse momentum of the produced hadron or jet, thus 
maintaining the clear-cut sensitivity to $\Delta g$ resulting from the 
direct piece. Focusing on the other hand on the resolved component, the study
of polarized photoproduction at HERA might even allow a measurement 
of the parton content of polarized {\em photons} in the long run \cite{sv} -- 
a unique task for HERA which makes the polarization upgrade option of HERA 
appear even more fascinating.

Polarized photoproduction reactions can also be studied in fixed target 
experiments with polarized lepton beam and polarized target, like the future 
COMPASS experiment at CERN, or HERMES at DESY. Among other things, 
one could look for charged tracks with large $p_T$ also in these 
experiments, whereas the energies would obviously not be 
large enough for producing decent jets. The resolved component at fixed target 
energies is expected to be generally negligible.

In order to make reliable quantitative predictions for a high-energy process
such as polarized inclusive-hadron photoproduction, it is crucial to extend
LO studies like the one of \cite{sv} to NLO by determining the ${\cal O}
(\alpha_s)$ QCD corrections. The key issue here is to check the perturbative 
stability of the process considered, i.e. to examine to what extent NLO 
corrections 
affect the cross sections and spin asymmetries relevant for experimental 
measurements. Only when the corrections are reasonably small and under control 
can a process that shows good sensitivity to, say, $\Delta g$ at the lowest 
order, be 
regarded as a genuine probe of the polarized gluon distribution and be 
reliably used 
to extract it from future data. The first basic ingredient for such an 
extension
to NLO has been provided in the past two years by the NLO fits to polarized 
DIS data mentioned above, which yielded spin-dependent nucleon
parton distributions evolved to NLO accuracy. Focusing on the direct part 
of inclusive-hadron photoproduction, the calculation of the polarized
cross section to NLO is then completed by using also (unpolarized) 
NLO fragmentation 
functions for the produced hadron (as provided in \cite{bkk}), and by 
including the ${\cal O} (\alpha_s)$ corrections to the 
spin-dependent ``direct'' subprocess cross
sections for the inclusive production of a certain parton that fragments into
the hadron. The calculation of the latter is the purpose of this paper.

An immediate problem arises here, as the direct part on its own is no longer 
a really well-defined quantity beyond the LO. 
This is due to the fact that beyond LO collinear singularities appear 
in the calculation of the subprocess cross sections for photon-parton 
scattering
which are to be attributed to a collinear splitting of the photon into a 
$q\bar{q}$ pair and need to be absorbed into the photon structure functions. 
As the latter only appear in the resolved part of the cross section, and since 
factorizing singularities is never a unique procedure, it follows that only 
the sum of the direct and the resolved pieces is independent of the 
factorization scheme 
chosen and thus is physical. This has been known for a long time from the 
unpolarized case where the corrections to the direct \cite{kr,lio} 
{\em and} to the resolved \cite{guil} contributions have all been calculated. 
Nevertheless, we will 
concentrate in this work only on the corrections to the direct part of the 
polarized cross section, mainly because this calculation -- albeit 
already being quite involved -- is much simpler than the one for the 
resolved piece. 
Our results will therefore only be the first step in a full calculation of NLO 
effects to polarized inclusive-hadron photoproduction. Despite the 
fact that they 
are not complete in the sense discussed above, we believe our 
results to be very 
important, both phenomenologically and theoretically: As mentioned earlier, 
the direct component dominates at fixed target energies and also still for 
the HERA 
collider situation in certain regions of phase space. This means that our 
NLO results should be rather close to the true NLO answer in these cases 
even if the 
resolved component is only taken into account on a LO basis, which in 
turn implies 
that our NLO corrections should already be sufficient to shed light on the 
question of 
general perturbative stability of the process. We also mention in this context
that our results for the NLO corrections to the direct hard subprocess 
cross sections will help to obtain or to check those for the resolved ones as 
the abelian (``QED-like'') parts of the two are the same. 

The paper is organized as follows: in sec.~2 we present the calculation of the
${\cal O} (\alpha_s)$ corrections to the direct part of polarized 
inclusive-parton 
photoproduction. Section 3 is devoted to a brief numerical evaluation 
of our results for HERA and fixed target kinematics. Section 4 contains the 
conclusions. 
\section{Calculation of the NLO corrections to the direct part of polarized 
inclusive-parton photoproduction}
\subsection{General framework}
The process we want to study is the single-inclusive production of a hadron $h$
in photoproduction in collisions of longitudinally polarized electrons 
(or muons) and protons, i.e. $\vec{e}(p_e) \vec{p}(p_p) \rightarrow 
h (p_h) X$. The NLO expression for the corresponding spin-dependent 
cross section is given by
\begin{eqnarray} \label{eq1}
E_h \frac{d\Delta\sigma^h}{d^3 p_h} &\equiv& 
\frac{1}{2} \Bigg( E_h \frac{d\sigma_{++}^h}{d^3 p_h} -
E_h \frac{d\sigma_{+-}^h}{d^3 p_h} \Bigg) \\
&=& \frac{1}{\pi S} \sum_{i,j} \int^1_{1-V+VW} 
\frac{dz}{z^2} \int_{VW/z}^{1-(1-V)/z} \frac{dv}{v(1-v)} 
\int^1_{VW/vz} \frac{dw}{w}  \nonumber \\ \label{eq2}
&& \Delta f_{\gamma}^e(x_e,M^2) \Delta f^p_i(x_p,M^2) D_j^{h}(z,M_F^2) \,
\frac{\pi\alpha_s(\mu^2)\alpha_{em}}{s} \, \times \\
&\times& \left[ 
\frac{d\Delta \hat{\sigma}^{(0)}_{\gamma i\rightarrow j}(v)}{dv} 
\delta (1-w) + \frac{\alpha_s(\mu^2)}{\pi} \frac{d\Delta 
\hat{\sigma}^{(1)}_{\gamma i\rightarrow j}}{dvdw}(s,v,w,\mu^2,M^2,M_F^2)
\right] \; , \nonumber
\end{eqnarray}
the subscripts ``$++$'', ``$+-$'' in~(\ref{eq1}) denoting the settings of the 
helicities of the incoming electron and proton. We have introduced the 
hadronic variables
\begin{eqnarray}
&& V\equiv 1+\frac{T}{S} \; \; , \;\;\;\; 
W\equiv \frac{-U}{S+T} \; \; , \nonumber \\
&&S\equiv (p_e+p_p)^2 \; \; , \;\;\;\; T\equiv (p_e-p_h)^2 \; \; , \;\;\;\; 
U\equiv (p_p-p_h)^2 \;\; ,
\end{eqnarray}
and the partonic ones 
\begin{eqnarray}
&& v\equiv 1+\frac{t}{s} \; \; , \;\;\;\; 
w\equiv \frac{-u}{s+t} \; \; , \nonumber \\
&&s\equiv (p_\gamma+p_i)^2 \; \; , \;\;\;\; t\equiv 
(p_\gamma-p_j)^2 \; \; , \;\;\;\; u\equiv (p_i-p_j)^2 \;\; .
\end{eqnarray}
Neglecting all masses, one has the relations
\begin{eqnarray}
&&s=x_e x_p S \; \; , \;\;\;\; t=\frac{x_e}{z} T \; \; , \;\;\;\; 
u=\frac{x_p}{z} U \; \; , \nonumber \\
&& x_e = \frac{VW}{vwz} \; \; , \;\;\;\; x_p = \frac{1-V}{z (1-v)} \; \; ,
\end{eqnarray} 
where $x_e$ ($x_p$) is the fraction of the longitudinal momentum of the 
electron (proton) taken by the photon (by parton $i$). Similarly, $z$ is 
the momentum share that hadron $h$ inherits from its parent parton $j$ in the
fragmentation process. The spin-dependent (``helicity-weighted'') parton 
distributions of electrons and protons that appear in the expression 
(\ref{eq2}) for the polarized cross section are defined as usual by
\begin{equation} \label{eqparton}
\Delta f_i^{e,p} (x,M^2) \equiv f_{i(+)}^{e,p(+)} (x,M^2) - 
f_{i(+)}^{e,p(-)} (x,M^2) \;\; ,
\end{equation}
where $f_{i(+)}^{e,p(+)} (x,M^2)$ ($f_{i(+)}^{e,p(-)} (x,M^2)$) denotes the
probability at scale $M$ of finding parton $i$ with positive helicity and 
momentum fraction $x$ in an electron or proton with positive (negative) 
helicity. As we only deal with the direct case, the only parton type 
occurring for the polarized electron structure functions is the photon, 
and the $\Delta f_\gamma^e$
coincide with the spin-dependent Weizs\"{a}cker-Williams 
spectrum\footnote{For the
resolved contribution, one has $\Delta f_\gamma^e \rightarrow 
\Delta f_k^e$ in (\ref{eq2}), 
where $\Delta f_k^e$ is a convolution of $\Delta f_\gamma^e$ in 
(\ref{eqweiz}) with the 
polarized photon structure function for parton type $k$.} \cite{ww}:
\begin{equation}  \label{eqweiz}
\Delta f_{\gamma}^e (y,M^2) \equiv \Delta P_{\gamma/e}(y) = 
\frac{\alpha_{em}}{2\pi} \left[ 
\frac{1-(1-y)^2}{y} \right] \ln \frac{Q^2_{max} (1-y)}{m_e^2 y^2} \; ,
\end{equation}
with $m_e$ being the electron (or muon) mass and $Q_{max}^2$ the 
allowed upper limit on the radiated photon's virtuality, to be fixed 
by the experimental conditions.
The fragmentation function $D_j^{h}(z,M_F^2)$ in (\ref{eq2}), describing the 
fragmentation process $j\rightarrow h$, is of course the usual 
unpolarized one since we sum over all polarizations in the final state.

Finally, the spin-dependent LO and NLO cross sections for the 
subprocesses $\gamma i
\rightarrow j X$, $d\Delta \hat{\sigma}^{(0)}_{\gamma i\rightarrow j}/dv$ and
$d\Delta \hat{\sigma}^{(1)}_{\gamma i\rightarrow j}/dvdw$, which have been 
stripped of trivial factors involving the electromagnetic coupling 
constant $\alpha_{em}$
and the strong one $\alpha_s(\mu^2)$, 
are defined in complete analogy with eq.~(\ref{eq1}). Note that, as indicated
in (\ref{eq2}), $d\Delta \hat{\sigma}^{(1)}_{\gamma i\rightarrow j}/dvdw$ will 
explicitly depend on the renormalization scale $\mu$ as a result of the 
renormalization procedure for the NLO virtual corrections, and also on the 
scales $M$, $M_F$ of the parton distributions and fragmentation functions, 
owing to the factorization of initial and final state collinear 
singularities. The calculation of the $d\Delta \hat{\sigma}^{(1)}_{\gamma 
i\rightarrow j}/dvdw$ is the purpose of this paper.

To conclude this section, let us note that the expression for the 
{\em un}polarized cross section for single-inclusive hadron 
photoproduction is similar to the one 
in eqs.~(\ref{eq1}),(\ref{eq2}), taking the sum instead of the difference in
(\ref{eq1}) and using unpolarized subprocess cross sections 
$d\hat{\sigma}^{(0)}_{\gamma i\rightarrow j}/dv$,
$d\hat{\sigma}^{(1)}_{\gamma i\rightarrow j}/dvdw$ and parton distributions 
in (\ref{eq2}). The latter correspond to taking the sum instead of the 
difference in (\ref{eqparton}), and for the electron case the unpolarized 
Weizs\"{a}cker-Williams spectrum \cite{ww} is obtained from (\ref{eqweiz}) by 
replacing $1-(1-y)^2 \rightarrow 1+(1-y)^2$. When calculating the 
polarized $d\Delta \hat{\sigma}^{(1)}_{\gamma i\rightarrow j}/dvdw$, we will
at the same time also determine their unpolarized counterparts and compare
them to existing analytical results in the literature~\cite{lio}. This will 
serve as a very good check on our calculation. Furthermore, the
 unpolarized cross 
section is needed when one wants to calculate spin asymmetries, defined by
\begin{equation}
A^h\equiv \frac{E_h d\Delta\sigma^h/d^3 p_h}{E_h d\sigma^h/d^3 p_h} \; \; ,
\end{equation}
which are usually the only quantities directly accessible to experiment.
\subsection{LO contributions}
The subprocesses contributing to $\Delta 
\hat{\sigma}^{(0)}_{\gamma i\rightarrow j}$ are 
\begin{eqnarray}
\vec{\gamma} \vec{q} &\rightarrow & g (q) \; \; , \nonumber \\
\vec{\gamma} \vec{q} &\rightarrow & q (g) \; \; , \nonumber \\
\vec{\gamma} \vec{g} &\rightarrow & q (\bar{q}) \; \; . \label{loproc}
\end{eqnarray}
Here it is understood that the final-state particle in brackets 
is unobserved and integrated over its entire phase space,
while the other fragments into the hadron. Note that the last process in 
(\ref{loproc}) is symmetric under exchange of $q$, $\bar{q}$. 
The corresponding spin-dependent cross sections read:
\begin{eqnarray}
\frac{d\Delta \hat{\sigma}^{(0)}_{\gamma q\rightarrow g}(v)}{dv}
&=& 2 C_F e_q^2 \frac{1-v^2}{v} \; \; , \nonumber \\
\frac{d\Delta \hat{\sigma}^{(0)}_{\gamma q\rightarrow q}(v)}{dv}
&=& 2 C_F e_q^2 \frac{1-(1-v)^2}{1-v} \; \; , \nonumber \\
\frac{d\Delta \hat{\sigma}^{(0)}_{\gamma g\rightarrow q}(v)}{dv}
&=& - 2 T_R e_q^2 \frac{v^2+(1-v)^2}{v(1-v)} \; \; , 
\end{eqnarray}
where $C_F=4/3$, $T_R=1/2$, and $e_q$ is the fractional charge of the quark. 
\subsection{NLO contributions}
Apart from the generic inclusive processes $\gamma q\rightarrow g$,   
$\gamma q\rightarrow q$, and $\gamma g\rightarrow q$ that are already present 
at the LO level, there are also contributions that can arise only beyond the 
Born approximation. These are $\gamma g\rightarrow g$, 
$\gamma q\rightarrow \bar{q}$,
and $\gamma q\rightarrow q'$, where in the latter process $q'$ denotes a 
quark (or antiquark) of flavour different from $q$. This means that the 
following explicit subprocesses have to be evaluated:
\begin{description}
\item[a)] the interference between the Born graphs 
$\vec{\gamma} \vec{q} \rightarrow g (q)$, $\vec{\gamma} \vec{q} \rightarrow q 
(g)$, $\vec{\gamma} \vec{g} \rightarrow q (\bar{q})$ and the virtual 
corrections to them,
\item[b)] the real corrections to the Born graphs, $\vec{\gamma} \vec{q} 
\rightarrow g (qg)$, $\vec{\gamma} \vec{g} \rightarrow q (\bar{q}g)$, and
$$\vec{\gamma} \vec{q} \rightarrow q \left\{
\begin{array}{l} 
(gg) \\
(q\bar{q}) \\ 
(q'\bar{q}') 
\end{array} \right. $$
(note that for the latter contribution a finite answer is obtained
only if all three subprocesses are added),
\item[c)] $\vec{\gamma} \vec{g} \rightarrow g (q\bar{q})$,
$\vec{\gamma} \vec{q} \rightarrow \bar{q} (qq)$,
$\vec{\gamma} \vec{q} \rightarrow q' (q\bar{q}') \; \; .$
\end{description}
\subsection{Regularization of singularities}
It is well known that singularities are encountered when calculating  
the loop diagrams or when performing the phase space 
integrations for the unobserved partons in the $2\rightarrow 3$ processes: 
first of all, the loop-diagrams contain ultraviolet divergencies 
which are removed by renormalization. Adding the renormalized loop and 
the corresponding $2\rightarrow 3$ contributions, the infrared singularities 
which are individually present in both parts, also cancel out, and one is 
left with collinear singularities which are finally removed by the 
factorization procedure (for the contributions from c) only 
singularities of the latter type occur). Of course, for being able to handle 
the singularities, one has to choose a consistent method of regularization.
In our calculation we use dimensional regularization for this purpose, where
$d=4-2 \epsilon$, which is the most convenient and customary 
choice. 

The calculation of the spin-dependent squared matrix 
elements requires projection onto definite helicity states of the incoming 
particles (which are taken to have momenta $p_1$, $p_2$), which is achieved by
using the relations 
\begin{equation} \label{qproj}
u(p_1,h_q) \bar{u}(p_1,h_q)=\frac{1}{2} \not \! p_1 (1-h_q \gamma_5 )
\end{equation}
for incoming quarks with helicity $h_q$ (analogously for antiquarks) and
\begin{equation} \label{gproj}
\epsilon^{\mu}(p_2,\lambda_g) \epsilon^{*\nu}(p_2,\lambda_g)=
\frac{1}{2(1-\epsilon)} \Bigg[ - g^{\mu \nu} + \frac{1}{p_1 \cdot p_2} 
\left( p_1^{\mu} p_2^{\nu} + p_1^{\nu} p_2^{\mu} \right) \Bigg] + 
\frac{i \lambda_g}{2 p_1\cdot p_2} \epsilon^{\mu \nu}_{\;\;\;\;\rho \sigma}
p_2^{\rho}p_1^{\sigma}
\end{equation}
for incoming gluons with helicity $\lambda_g$. The parts independent of $h_q$ 
and $\lambda_g$ contribute to the {\em un}polarized matrix elements, for which 
the averaging of gluon spins in $d$ dimensions should be performed by dividing 
by the $d-2=2 (1-\epsilon)$ possible spin orientations, as has been 
made explicit in eq.~(\ref{gproj}). 

As is well known, the use of $\gamma_5$ and the Levi-Civita tensor 
appearing in 
(\ref{qproj}),(\ref{gproj}) is not entirely straightforward in 
$d\neq 4$ dimensions.
For our calculations we will use the original prescription of `t Hooft and 
Veltman \cite{tv}, afterwards systematized by Breitenlohner and Maison 
\cite{bm} 
(HVBM scheme), which is usually regarded as the most reliable scheme in the 
sense that its internal algebraic consistency is well established.
In this scheme explicit definitions for $\gamma_5$ and 
$\epsilon_{\mu\nu\rho\sigma}$ are given. In particular, 
$\gamma_5 \equiv i\epsilon^{\mu\nu\rho\sigma} \gamma_{\mu}
\gamma_{\nu}\gamma_{\rho}\gamma_{\sigma}/4!$, the $\epsilon$-tensor 
being regarded 
as a genuinely four-dimensional object with its components vanishing in all 
unphysical dimensions. In this way the $d$-dimensional Minkowski space is
explicitly divided into two subspaces, a four-dimensional one and a
$(d-4)$-dimensional one, each of them equipped with its metric tensor. 
As a result, apart from $d$-dimensional scalar products $p\cdot q$ (the usual 
Mandelstam variables), also their respective ``subspace'' counterparts 
can show up in calculations, which renders the calculation of traces and 
phase space integrations somewhat more complicated. Fortunately, we can rely
in our calculation to a certain extent on known results, as will be 
discussed in the next subsection. 
\subsection{Virtual corrections and $2\rightarrow 3$ matrix elements}
In \cite{wv} the NLO corrections to the (``non-fragmentation'') 
part of the hadronic 
single-spin cross section for the production of {\em circularly polarized} 
prompt photons, i.e. the QCD corrections for $\vec{p}p\rightarrow 
\vec{\gamma}X$, were calculated. This calculation 
involved the virtual corrections to the Born graphs $\vec{q}g\rightarrow 
\vec{\gamma}q$, $\vec{g}q\rightarrow \vec{\gamma}q$, 
$\vec{q}\bar{q}\rightarrow \vec{\gamma}g$, as well as the 
$2\rightarrow 3$ matrix elements $\vec{a}b\rightarrow
\vec{\gamma}cd$. These ingredients were obtained in \cite{wv} 
in the HVBM scheme.
We therefore can get the virtual corrections for $\vec{\gamma} \vec{q} 
\rightarrow g q$, $\vec{\gamma} \vec{q} \rightarrow q g$, 
$\vec{\gamma} \vec{g} \rightarrow q \bar{q}$ and the $2\rightarrow 3$ 
cross sections $\vec{\gamma}\vec{a} \rightarrow bcd$ by appropriately 
crossing the polarized photon with the unpolarized incoming parton 
in the results of \cite{wv}, which greatly facilitates the calculation. 

The virtual corrections obtained in this way read in the 
$\overline{{\rm MS}}$ scheme:
\begin{eqnarray}
\lefteqn{  \!\!\!
\frac{d\Delta\hat{\sigma}_{q\gamma \rightarrow q}^{(1),V}}{dvdw}=
\frac{C_F \, e_q^2 \mu^{2\epsilon}}{\Gamma (1-2\epsilon)}
\left(
\frac{(4\pi\mu^2)^2}{s^2 v (1-v)}\right)^{\epsilon}\, \delta (1-w)
\left[
   - \frac{2\,C_F + N_C}{\epsilon^2} \,\delta T_{q\gamma}  
\right.} \nonumber \\  
&-& \frac{1}{\epsilon} 
\left(
 b_0\, \delta T_{q\gamma}-  2\,C_F\,\ln v_1\, \delta T_{q\gamma}+ 
         N_C \, \delta T_{q\gamma}\ln \frac{v_1}{v} \,
          +   N_C\, \frac{v_1^2}{v} +
  C_F \,\frac{v_1}{v} \, \left( 5 + v \right) \right)
  \nonumber \\
 &+& b_0\,\ln \frac{\mu^2}{s}\,\delta T_{q\gamma}
 -  \left( 2\,C_F - N_C \right) \,
 \ln v_1\,\ln v\,      
      \frac{1 - 2\,v}{v} + \left( 2\,C_F - N_C
         \right) \,
  \ln^2 v\,\frac{1 - 2\,v}{2\,v} \nonumber   \\  
&+&  C_F\,\ln v\,\frac{3 - 2\,v}{v} - 
  b_0\, \frac{ {v_1}^2}{v} 
+ \left( 2\,C_F - N_C \right) \,
  \frac{{v_1^2}}{v} \, \ln v_1\,  
- 2\,C_F\, v_1 \,
      \frac{5 + 2\,v}{v} \nonumber   \\  
  &+& \left. C_F\,\pi^2\,\frac{2 - 6\,v + v^2}
    {3\,v} + N_C\,\ln v\,
      \frac{1 - v + v^2}{v} - 
  N_C\, \pi^2 \,\frac{1 - 6\,v + 2\,{v^2}}
    {6\,v}   \right] \:\:\: ,
\end{eqnarray}
\begin{equation}
\hspace*{-7.3cm}
\frac{d\Delta\hat{\sigma}_{q\gamma \rightarrow g}^{(1),V}}{dvdw}=
\frac{d\Delta\hat{\sigma}_{q\gamma \rightarrow q}^{(1),V}}{dvdw}
\Big[ v \longleftrightarrow (1-v) \Big] \;\;\; ,
\end{equation}
\begin{eqnarray}
\lefteqn{ \!\!\!\!\!\!\!  
\frac{d\Delta\hat{\sigma}_{g\gamma \rightarrow q}^{(1),V}}{dvdw}=
\frac{T_R e_q^2 \mu^{2\epsilon}}{\Gamma (1-2\epsilon)}
\left(
\frac{(4\pi\mu^2)^2}{s^2 v (1-v)}\right)^{\epsilon}  
\delta (1-w)  \left[   
- \frac{2\,C_F + N_C}{\epsilon^2} \,
      \delta T_{g\gamma}  \right. }\nonumber \\
 &-& \frac{1}{\epsilon} \left( b_0\,\delta T_{g\gamma}
         + 3\,C_F\,\delta T_{g\gamma} - 
              N_C\,\delta T_{g\gamma}\, \ln (v v_1) \right)
 +N_C \ln (v v_1) \nonumber \\ 
& -& 7\,C_F\,\delta T_{g\gamma} 
 + b_0 \,\ln \frac{\mu^2}{s}\,\delta T_{g\gamma} - 
  N_C\,\delta T_{g\gamma} \ln v_1\,\ln v + 
  \frac{1}{6}\left( 4\,C_F - N_C \right) \,{{\pi }^2}\,
      \delta T_{g\gamma} \nonumber \\
  &-& 
  C_F\,\ln v\, \frac{3 - v}{v} 
- C_F\,\ln v_1\,\frac{2 + v}{1 - v} 
-   \left( 2\,C_F - N_C \right) \, \ln^2 v \,\frac{1 + {v^2}}
{2\,\left( 1 - v \right) \,v} \nonumber \\ 
&-& \left. 
      \left( 2\,C_F - N_C \right) \, \ln^2 v_1\,
      \frac{2 - 2\,v + {v^2}}{2\,\left( 1 - v \right) \,v} 
   \right] \:\:\: ,
\end{eqnarray}
where $N_C=3$, $b_0=11 N_C/6 -n_f/3$ ($n_f$ being the number of 
active flavours), and $\mu$ is the renormalization scale. Furthermore, 
\begin{eqnarray}
\delta T_{q\gamma} &=& (1-v^2)/v \:\:\: , \\
\delta T_{g\gamma} &=& -v_1/v-v/v_1 \:\:\: , 
\end{eqnarray}
with $v_1=1-v$. Note that the result for $\vec{q} \vec{\gamma} 
\rightarrow qg$ can also be obtained from the one of \cite{gv} 
for $\vec{q}\vec{g} \rightarrow \gamma q$ after crossing and correct 
adjustment of colour.

The integration of the real $2\rightarrow 3$ matrix elements over the
phase space of the unobserved particles has been discussed in detail 
in \cite{effh,gv} and needs not be recalled here. The technical 
complications related to the use of the HVBM scheme discussed above 
have been solved in \cite{gv}.

Adding the renormalized virtual and the real contributions, all infrared
singularities cancel out. In the next section we briefly recall the 
factorization procedure which removes the remaining collinear singularities.
\subsection{Factorization}
The factorization procedure based on the factorization theorem 
\cite{fac} has been outlined in, for instance, refs. \cite{effh,gv}. 
The mass singularities arise when either an incoming particle
collinearly emits another particle (or splits into a pair of 
collinear particles), or when the ``observed'' final state particle  
is collinear to an unobserved one. The singular terms attached to the 
initial legs are separated off at the factorization scale $M$ and 
absorbed into the initial-state 
parton distributions which then obey NLO QCD evolution equations. 
In particular, 
if the singularity results from a collinear splitting $\gamma 
\rightarrow q\bar{q}$, 
it is absorbed into the ``pointlike'' part of the photon structure function. 
Of course there is freedom in choosing the factorization prescription, i.e. in
subtracting finite pieces along with the pole terms. As already pointed 
out in the 
introduction, this is the reason why a separation of direct and resolved 
contributions 
to a photoproduction cross section becomes, strictly speaking, 
meaningless beyond LO. 
Final state singularities are factorized at the scale $M_F$ into the 
(NLO) unpolarized fragmentation functions $D_f^h$. 

As an example, let us briefly discuss the factorization of the polarized 
$\vec{q} \vec{\gamma} \rightarrow g(qg)$ subprocess. This is performed in
the easiest way by adding a ``counter cross section'' \cite{effh} which, 
taking into consideration all possible collinear configurations, has the form
\begin{eqnarray}
\frac{d\Delta \hat{\sigma}_{q\gamma \rightarrow g}^{(1),F}}{dvdw} 
&\sim&-\frac{\alpha_s}{2\pi} \left[  \int_0^1 dx_1 \Delta H_{qq}(x_1,M^2)
\frac{d\Delta \hat{\sigma}_{\epsilon}^{q\gamma\rightarrow gq}}{dv}(x_1 s,1+
\frac{t}{s}) \;
\delta \left( x_1 (s+t)+u \right) \right. \nonumber \\
&+&\int_0^1 \frac{dx_3}{x_3^2} H_{gg} (x_3,M_F^{2}) 
\frac{d\Delta \hat{\sigma}_{\epsilon}^{q\gamma\rightarrow gq}}{dv}(s,
1+\frac{t}{x_3 s}) \; \delta \left( s +\frac{1}{x_3} (t+u) \right) \nonumber 
\\ &+&\left. \int_0^1 \frac{dx_3}{x_3^2} H_{gq} (x_3,M_F^{2}) 
\frac{d\Delta \hat{\sigma}_{\epsilon}^{q\gamma\rightarrow qg}}{dv}(s,
1+\frac{t}{x_3 s}) \; \delta \left( s +\frac{1}{x_3} (t+u) \right) \right] 
\nonumber \\ &&\hspace*{-1.8cm}
-\frac{\alpha_{em}}{2\pi}  \left[ \int_0^1 dx_2 \Delta H_{q\gamma}(x_2,M^2)
\frac{d\Delta \hat{\sigma}_{\epsilon}^{q\bar{q}\rightarrow gg}}{dv}(x_2 s,
1+\frac{t}{x_2 s})
\; \delta \left( x_2 (s+u)+t \right) \right] ,
\end{eqnarray}
where the $d\Delta\hat{\sigma}_{\epsilon}^{ab\rightarrow cd}(s,v) /dv$ are
the polarized $d$-{\em dimensional} $2\rightarrow 2$ cross sections 
for the processes $ab\rightarrow cd$, to be found for the HVBM scheme 
in \cite{gv}. Furthermore,
\begin{equation} \label{hdef}
(\Delta) H_{ij} (z,M^2) \equiv -\frac{1}{\hat{\epsilon}} (\Delta)
P_{ij} (z) \left( \frac{\mu^2}{M^2} \right)^{\epsilon} +
(\Delta) f_{ij} (z)  \; ,
\end{equation}
where $1/\hat{\epsilon}\equiv 1/\epsilon-\gamma_E+\ln 4\pi $, as usual
in the $\overline{{\rm MS}}$ scheme. In eq.~(\ref{hdef}) the
$(\Delta) P_{ij}(z)$ denote the unpolarized (polarized) one-loop
splitting functions for the transitions $j\rightarrow i$ \cite{ap}. The
functions $(\Delta) f_{ij}(z)$ represent the freedom in choosing a 
factorization prescription. In the $\overline{{\rm MS}}$ scheme these 
functions vanish. Note that even in the polarized case only the 
{\em un}polarized $H_{ij}$ contribute to the factorization of final-state 
singularities, as we do not consider the production of polarized hadrons.
 
Before proceeding, we have to mention an important subtlety related to the 
use of the HVBM prescription for $\gamma_5$, which affects the polarized 
function $\Delta H_{qq}$. It is a well-known property of the 
HVBM--$\gamma_5$ that
it leads to helicity non-conservation at the $qqg$ vertex in $d$ dimensions, 
expressed by a non-vanishing difference of unpolarized and polarized
$d$-dimensional LO quark-to-quark splitting functions,
\begin{equation} \label{pqqdim}
\Delta P_{qq}^{4-2 \epsilon} (x)-P_{qq}^{4-2 \epsilon} (x) =
4 C_F \epsilon (1-x) \; .
\end{equation}
A disagreeable consequence of this is a non-zero first moment 
($x$-integral) of the non-singlet NLO anomalous dimension for the evolution of 
polarized non-singlet quark densities, in obvious conflict with the 
conservation of the flavour non-singlet axial current \cite{svw,vn,wv1}. 
At the same time, (\ref{pqqdim}) is responsible for producing a result 
for the ${\cal O}(\alpha_s)$ correction to the Bj\o rken sum rule 
\cite{bjsr} which disagrees with the one of \cite{rat}.
These two effects turn out to be closely related, as they can be simultaneously
removed by a factorization scheme transformation \cite{vn,wv1}, generated by 
the term on the right-hand-side of (\ref{pqqdim}). In other words, 
it is advisable,
albeit not mandatory in a purely mathematical sense, to slightly deviate from
the $\overline{\rm{MS}}$ scheme in the polarized case by choosing (see also
\cite{wv1}) 
\begin{equation}
\Delta f_{qq} (z) = -4 C_F (1-z)
\end{equation}
in (\ref{hdef}).
The factorization scheme transformation defined by this equation has also been 
performed in the calculations of the spin-dependent NLO splitting functions 
\cite{vn,wv1} and is thus respected by the available sets of spin-dependent NLO
parton densities [1--4]. The ``$\gamma_5$-effect'' described 
above has been known to occur in the HVBM scheme for quite some time
\cite{alex,arg2,svw,kamal} and is obviously a pure artefact of the 
regularization prescription chosen. Since furthermore physical 
requirements such as the conservation of the non-singlet axial current 
serve to remove the effect in a straightforward and obvious way, results 
of NLO calculations in ``spin-physics'' 
(like the ones of \cite{vn,wv1}, or ours) are usually regarded as 
being ``genuinely'' in the conventional $\overline{\rm{MS}}$ 
scheme only {\em after} this transformation has been carried out. 
The quantities $\Delta f_{qg}$, $\Delta f_{gq}$, and 
$\Delta f_{gg}$ in (\ref{hdef}) will of course be set to zero, as in the 
usual $\overline{\rm{MS}}$ scheme. Needless to say that in the unpolarized 
case ($\overline{\rm{MS}}$) one has $f_{qq}=f_{qg}=f_{gq}=f_{gg}=0$.  

Another comment concerns the functions $(\Delta) H_{q\gamma}$ needed for 
factorizing initial-state collinear singularities from photon-splitting to a 
$q\bar{q}$ pair. As mentioned above, such singularities are absorbed into 
the ``pointlike'' part of the photon structure functions. Studies 
\cite{grvgam,aur,svgam} of the photon structure beyond LO have revealed 
that the $\overline{\rm{MS}}$-scheme photonic coefficient functions for the
photon's DIS structure functions $F_2^{\gamma}$, $g_1^{\gamma}$ exhibit
a logarithmically singular behaviour at large $x$. Combining at NLO 
the ``pointlike'' 
parts of $F_2^{\gamma}$, $g_1^{\gamma}$ with estimates for the ``hadronic''
component based on vector meson dominance (VMD) arguments, one encounters
strongly negative results at large $x$, ruling out the use of intuitive 
VMD ideas in the $\overline{\rm{MS}}$ scheme. Instead, an appropriately 
adjusted (``fine tuned'') non-VMD hadronic NLO input would be required in 
the $\overline{\rm{MS}}$ scheme, substantially differing from the 
LO one, as the only means of avoiding unwanted and physically not acceptable 
perturbative instabilities for physical quantities like $F_2^{\gamma}$, 
$g_1^{\gamma}$. In the unpolarized case the so-called $\rm{DIS}_{\gamma}$ 
factorization scheme \cite{grvgam} was introduced to avoid such 
``inconsistencies''. 
Here the idea was to absorb the photonic Wilson coefficient for $F_2^{\gamma}$ 
into the photon's quark densities by a factorization scheme transformation,
hereby leaving the ``hadronic'' part untouched. In \cite{svgam}, 
this procedure was 
extended to the polarized case. It was found that after transforming to 
the $\rm{DIS}_{\gamma}$ scheme, a pure VMD input can be successfully used 
for phenomenological analyses going beyond the LO.
We will therefore specify the functions $(\Delta) f_{q\gamma}$ to be used to 
transform to the $\rm{DIS}_{\gamma}$ scheme. They read:
\begin{eqnarray}
f_{q\gamma} (x) &=& T_R \left[ (x^2 + (1-x)^2)  \left( \ln \frac{1-x}{x} -1 
\right)
+6 x (1-x) \right] \;\; , \nonumber \\
\Delta f_{q\gamma} (x) &=& T_R \left[ (2x-1) \left( \ln \frac{1-x}{x} -1 
\right) +
2 (1-x) \right] \;\; ,
\end{eqnarray}
where $T_R=1/2$. Of course, the choice of factorization 
scheme cannot affect the result for a physical quantity. In other words, in the
unpolarized case, where all contributions can be consistently calculated to
NLO, it does not matter eventually whether we use photonic parton densities 
defined in the $\rm{DIS}_{\gamma}$ or the $\overline{\rm{MS}}$ scheme,
as long as we use NLO hard cross sections determined in the same scheme. 
In the polarized case however, we are not yet able to consistently include 
the NLO ``resolved'' contributions, as was pointed out several times before. 
Therefore, comparing the results for the direct part of the
NLO cross section in the $\overline{\rm{MS}}$ and 
the $\rm{DIS}_{\gamma}$ schemes might indicate the uncertainty 
resulting from not performing a consistent NLO calculation.
\subsection{Final results}
For all processes the final partonic cross section can be cast into the form:
\begin{eqnarray}
\frac{d\Delta \hat{\sigma}_{\gamma i\rightarrow j}^{(1)}}{dvdw} 
(s,v,w,\mu^2,M^2,M_F^2) &=&
\left[ \left(  c_a  \delta (1-w) +   c_b 
\frac{1}{(1-w)_+} +  c_c \right) \ln \frac{M^2}{s} \right. \nonumber \\
&+&\left(  c_{\tilde{a}}  \delta (1-w) +   c_{\tilde{b}} 
\frac{1}{(1-w)_+} +  c_{\tilde{c}} \right) \ln \frac{M_F^2}{s} \nonumber \\
&+&\tilde{c}_1 \delta (1-w) \ln \frac{\mu^2}{s} + c_1 \delta (1-w)  
\nonumber \\
&+& c_2 \frac{1}{(1-w)_+} +  c_3 \left(
\frac{\ln (1-w)}{1-w} \right)_+ + c_4 \ln v \nonumber  \\
&+& c_5 \ln (1-v) +   c_6 \ln w+  c_7 \frac{\ln w}{1-w}
+  c_8 \ln(1-w) \nonumber \\
&+&   c_9 \ln (1-vw) +   c_{10} \frac{\ln \frac{1-v}{1-vw}}{1-w}+
  c_{11} \ln (1-v+vw)  \nonumber \\
&+& \left.   c_{12} \frac{\ln (1-v+vw)}{1-w} +  c_{13} \right] \:\:\: .
\label{final}
\end{eqnarray}
Distributions in $w$ like $\delta (1-w)$, $1/(1-w)_+$, etc. only occur
for the subprocesses that are already present at the Born level. An expression 
similar to (\ref{final}) holds for the unpolarized case with, obviously,
different coefficients $c_i (v,w)$. We note that we have compared our 
unpolarized results to the ones presented in an analytical form in 
\cite{lio}. We found an almost complete overall agreement; however, there 
are a very small number of differences, some of which could be related to 
typographical mistakes. The only major discrepancies arise for the 
subprocesses $\gamma q \rightarrow q (q' \bar{q}')$ and $\gamma 
q\rightarrow q' (q \bar{q}')$. For the first, we believe that the result 
in \cite{lio} was accidentally presented in terms of 
the ``crossed'' process $q \gamma \rightarrow q (q' \bar{q}')$.
For $\gamma q\rightarrow q' (q \bar{q}')$, it seems that the result 
in \cite{lio} rather corresponds to $\gamma q\rightarrow \bar{q}' (q q')$. 
Anyway, none of these small discrepancies turns out to have a 
significant numerical effect. The coefficients $c_i (v,w)$ for the 
unpolarized and polarized cases are rather lengthy and will not be given 
here. They can be obtained in a Fortran code via electronic
mail from Werner.Vogelsang@cern.ch.
\section{Numerical results}
Let us now present some first numerical results for the NLO corrections 
to polarized single-inclusive photoproduction of charged hadrons. 
Rather than performing a detailed numerical study of the process, 
we will restrict ourselves to the most interesting questions. 
These concern the general size of the corrections (``$K$-factors'') 
and the residual dependence of the NLO cross section on the unphysical 
scales present in the calculation.

Before starting, we mention that whenever we will calculate the unpolarized
NLO cross section, we will do so in a completely consistent way, 
i.e. by including 
both the direct {\em and} the resolved parts at NLO. Here we make use of 
our own results for the NLO corrections to the direct part of the cross section
(see sec.~2), and of the ones in \cite{guil} for the NLO resolved part. 
Furthermore, we will for consistency use NLO parton densities for 
the incoming proton \cite{grv} and the photon \cite{grvgam}, as well as 
NLO fragmentation functions. For the latter we will use the ones of 
\cite{bkk} set up for the sum of charged pions and kaons. 
They will also be our choice when calculating the polarized cross section.

In the polarized case at NLO, we will use spin-dependent parton 
distributions for the proton evolved at NLO and fitted to the available 
DIS data. Several sets for these are available \cite{grsv,gs,daniel}; 
for definiteness we will choose the ones of \cite{grsv} 
determined within the ``radiative parton model''. These also have the 
agreeable property of providing parametrizations at NLO and LO, 
the latter to be used for Born level predictions. In particular, we will 
choose the ``valence'' set of \cite{grsv}, which corresponds to the 
best-fit result of that paper, along with one other set of \cite{grsv} 
based on assuming $\Delta g (x,\mu^2) = g(x,\mu^2)$ at the low input scale 
$\mu$ of \cite{grsv}, where $g(x,\mu^2)$ is the unpolarized GRV \cite{grv} 
input gluon distribution. This set will be referred to as ``max. gluon''
in what follows. Employing these two sets, which both provide a good 
fit to the available DIS data, but differ significantly in the polarized 
gluon density, we are able to see to which extent the relative size of the 
NLO corrections depends on the set of parton distributions used.

We also note that whenever we calculate a cross section at LO (for instance,
when determining the $K$-factor $K=\sigma^{NLO}/\sigma^{LO}$), we will 
for consistency use LO parton distributions and fragmentation functions. 
In this case we will also use the one-loop expression for the strong 
coupling, whereas at NLO we obviously employ its two-loop counterpart. 
The LO/NLO values for the QCD scale parameter $\Lambda_{QCD}^{(n_f)}$ 
for $n_f$ active flavours are taken from \cite{grv,grvgam,grsv}.
Heavy flavour ($c$, $b$) contributions to the cross sections are neglected for
simplicity. Unless we explicitly study the scale dependence of our results,
we will choose the renormalization and factorization scales to be equal to the
transverse momentum $p_T$ of the produced hadron.  

We will provide numerical results for both the fixed target and the HERA 
collider kinematic domains. While the resolved component is expected to be 
generally small at fixed target energies, it is known \cite{sv} to be 
dominant in certain regions of phase space at HERA also for the polarized 
case. Here the direct contribution will dominate only 
%
\begin{figure}[t]
\begin{center}
\vspace*{-1.4cm}
\epsfig{file=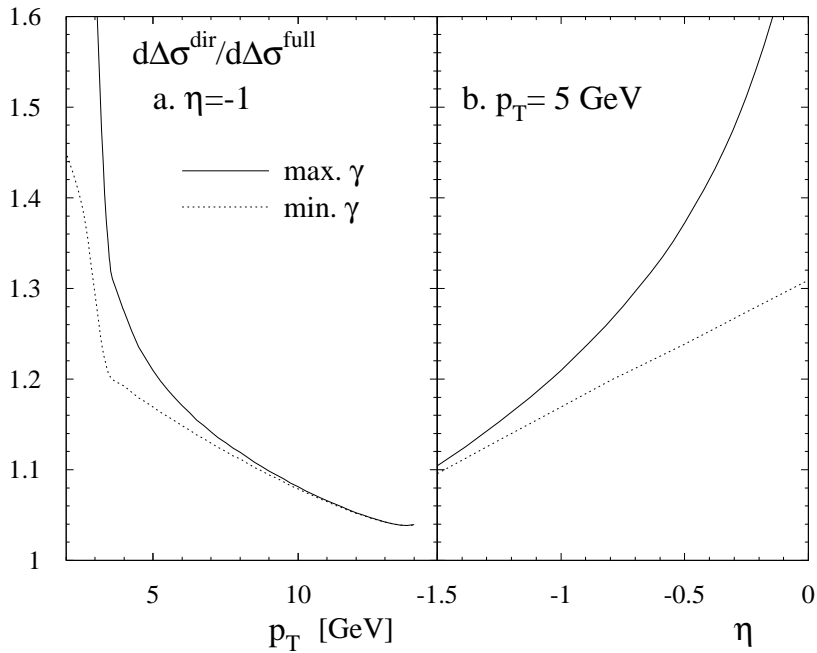,width=12.5cm}
\vspace*{-0.1cm}
\caption{\sf Ratio of LO direct and full (direct + resolved) 
polarized cross sections
for HERA energies ($E_e=27$ GeV, $E_p=820$ GeV). {\bf a.} $p_T$ dependence at  
$\eta=-1$, {\bf b.} $\eta$ dependence at $p_T=5$ GeV.}
\vspace*{-0.65cm}
\end{center}
\end{figure} 
at fairly large $p_T$, and/or at negative rapidities
$\eta$ of the produced hadron in the HERA laboratory frame, where we have, 
as usual, counted positive rapidity in the proton forward direction. 
In order to demonstrate this, and to isolate for our further
HERA studies the region where the direct contribution dominates, 
figure~1 shows 
the ratio of the direct part of the polarized cross section over the full 
(direct + resolved) one, calculated at LO and plotted vs. $p_T$ (at $\eta=-1$) 
and $\eta$ (at $p_T=5$ GeV). We have assumed $E_e=27$ GeV and 
$E_p=820$ GeV; the cuts 
on the polarized Weizs\"{a}cker-Williams spectrum were chosen as in \cite{sv}. 
We have used the GRSV ``max. gluon'' set for the polarized proton. 
For the LO resolved part in the denominator we have to pick a suitable set of
LO parton distributions for the polarized photon. Of course, nothing is 
known as yet experimentally about the latter, so we need to resort to 
models for them. Here we will follow \cite{sv} to use two very different 
scenarios, first considered in \cite{gvg}. They are based on assuming 
``maximal'' ($\Delta f^{\gamma}(x,\mu^2)=f^{\gamma}(x,\mu^2)$) or ``minimal'' 
($\Delta f^{\gamma}(x,\mu^2)=0$) saturation of the fundamental positivity 
constraints $|\Delta f^{\gamma}(x,\mu^2)| \leq f^{\gamma}(x,\mu^2)$ at the
input scale $\mu$ for the QCD evolution, where $\mu$ and the unpolarized 
photon structure functions $f^{\gamma}(x,\mu^2)$ were adopted from the 
phenomenologically successful radiative parton model predictions in 
\cite{grvgam}. These sets will be dubbed ``max. $\gamma$'' 
and ``min. $\gamma$'' sets, respectively, and figure~1 shows the results 
obtained for both sets. As can be seen, in the region defined by 
$\eta \leq -1$, $p_T\geq 5$ GeV the resolved component is expected to 
contribute about 20$\%$ or less to the cross section (note that the direct 
and resolved parts of the cross section turn out to be of opposite sign).

Having determined the region where the direct component dominates for 
HERA energies, we can now turn to NLO. Figure~2 shows the $K$-factors for 
the direct part of the polarized cross section in the $\overline{{\rm MS}}$ 
scheme, again vs. $p_T$ (at $\eta=-1$) and $\eta$ (at $p_T=5$ GeV). 
The solid line corresponds to the ``max. gluon'' set for the polarized 
parton densities of the proton, whereas the dashed one displays
the result obtained within the ``valence'' best-fit scenario of \cite{grsv}. 
As one can clearly see, the $K$-factors are of very moderate size, 
$K\lesssim 1$ for almost all $p_T$ and $\eta$ examined. Only at very large 
$p_T$, near the edge of phase space for the $\eta=-1$ considered, does the 
$K$-factor become much larger than unity within the ``valence'' scenario. 
This finding of generally small NLO corrections is very important and 
corroborates the LO predictions previously made in \cite{sv}.
%
%
\begin{figure}[ht]
\begin{center}
\vspace*{-1.4cm}
\epsfig{file=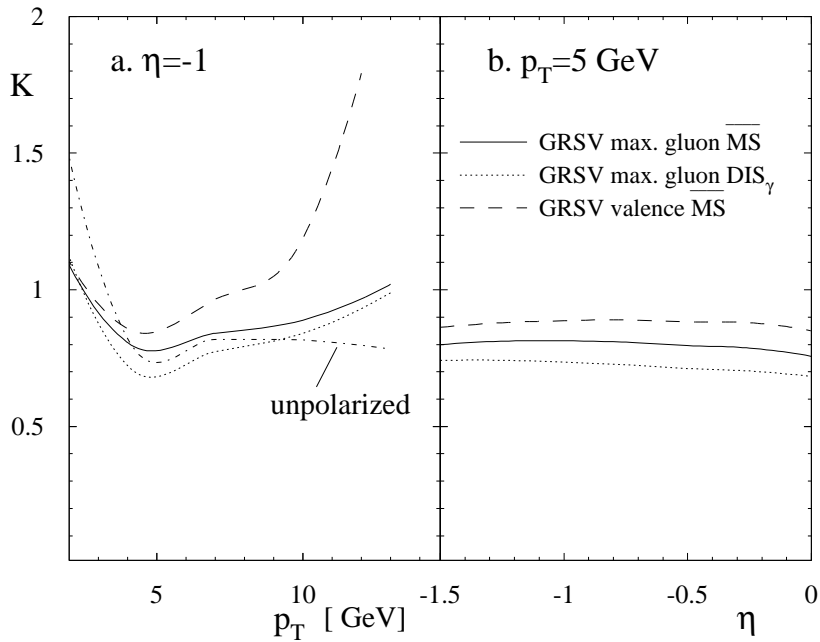,width=12.5cm}
\vspace*{-0.1cm}
\caption{\sf $K$-factors for the direct part of the polarized cross 
section at HERA energies for different GRSV \cite{grsv} parton distributions. 
In {\bf a.} the K-factor for the total unpolarized cross section is also 
shown.}
\vspace*{-0.65cm}
\end{center}
\end{figure} 

As frequently mentioned earlier, the NLO direct part on its own is 
factorization scheme dependent. For comparison we also plot in 
fig.~2 the $K$-factor for the direct cross section obtained within the 
$\rm{DIS}_{\gamma}$ scheme introduced in sec.~2.6. As 
can be seen, the corresponding change of the result is rather small. Finally,
figure~2 also presents the $K$-factor for the {\em full} (direct + 
resolved) unpolarized cross section, which of course is scheme-independent. 
It turns out that it is very similar in size and shape to the 
$K$-factors we have obtained for the direct part of the polarized 
cross section. This, again, is a very satisfactory finding, as
it suggests that our $K$-factor for the direct part might not be too far off
the result for the one of the full polarized cross section, to be eventually
determined when the NLO corrections to the resolved part of the 
polarized cross section will have been calculated.

%
%
\begin{figure}[ht]
\begin{center}
\vspace*{-0.65cm}
\epsfig{file=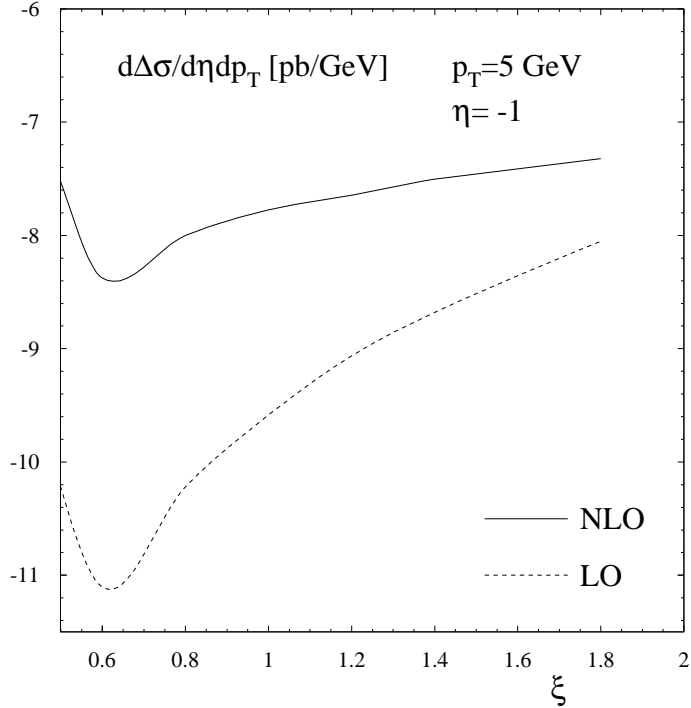,width=10.5cm}
\vspace*{-0.1cm}
\caption{\sf Scale dependence of the direct part of the polarized 
inclusive-hadron photoproduction cross section at $p_T=5$ GeV and 
$\eta=-1$ for HERA energies. All scales have been set equal to 
$\sqrt{\xi} p_T$, and the parton distributions used correspond to the 
``max. gluon'' set of \cite{grsv}. The NLO cross section has been 
calculated in the $\overline{{\rm MS}}$ scheme.}
\vspace*{-0.65cm}
\end{center}
\end{figure} 
Another important issue when going beyond the LO is the expected 
reduction in the dependence of the results on the unphysical scales 
$\mu$, $M$, $M_F$ introduced previously. We now set 
$\mu^2=M^2=M_F^2=\xi p_T^2$ and plot in fig.~3 the LO and NLO direct 
cross sections as functions of $\xi$ for fixed $\eta=-1$, $p_T=5$ GeV. 
Even though we can only consider the direct part, the improvement in the 
scale dependence when going from LO to NLO becomes already clearly visible. 

We finally turn to the fixed target region, relevant for the HERMES and the 
future COMPASS experiments. It is again interesting to study the size of the 
$K$-factor for this situation, choosing a muon beam energy of 200 GeV.
The results for our two sets of polarized parton densities of the proton are 
displayed as functions of $p_T$ in fig.~4, where we have fixed the 
centre-of-mass rapidity, $\eta_{cm}=0$. We have again calculated the 
NLO cross section in the $\overline{{\rm MS}}$ scheme. One can clearly see 
that again the $K$-factors are of very reasonable size, once $p_T \geq 3$ GeV, 
where one intuitively would start to trust perturbation theory.
%
\begin{figure}[ht]
\begin{center}
\vspace*{-1.4cm}
\epsfig{file=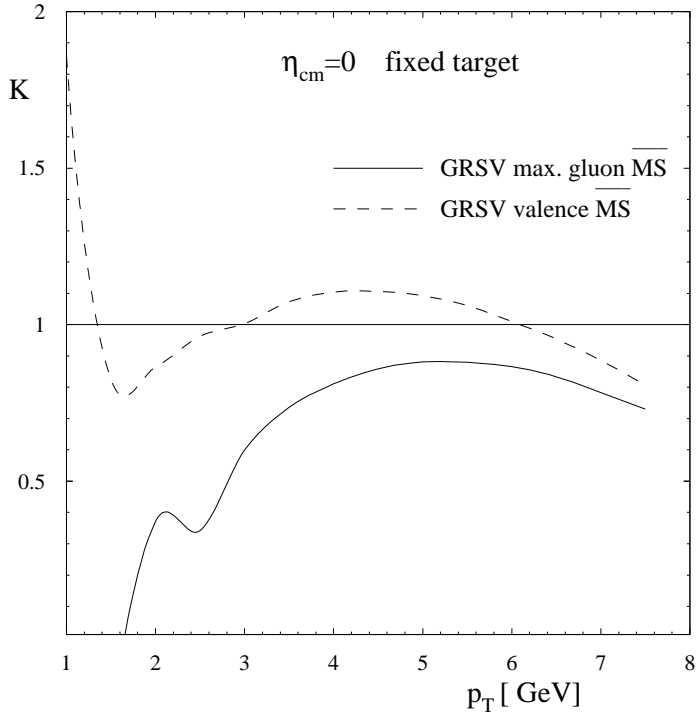,width=10.5cm}
\vspace*{-0.1cm}
\caption{\sf $K$-factors for the direct part of the polarized single-inclusive 
charged-hadron cross section in a fixed target experiment with $s=400$ GeV$^2$ 
at $\eta_{cm}=0$.}
\end{center}
\end{figure} 

\vspace*{-0.2cm}
\section{Summary and Conclusions}
\vspace*{-0.3cm}
We have presented for the first time the next-to-leading order QCD corrections 
to the spin-dependent cross section for single-inclusive charged-hadron 
photoproduction. This process derives its importance from its sensitivity to 
the proton's spin-dependent gluon distribution and, at high energies, to 
the so far completely unknown parton content of circularly polarized 
quasi-real photons. It could be studied experimentally in future polarized 
fixed-target lepton-nucleon experiments, but also at the HERA $ep$ collider 
after an upgrade to both beams being polarized. 

Our calculation is an important first step in trying to assess the 
perturbative stability of this process. First numerical results show 
generally moderate NLO corrections for the direct part of the cross section, 
the $K$-factor being close to unity over a wide kinematical range at both 
HERA and fixed target energies. Also, the expected reduction in scale 
dependence of the cross section when going from LO to NLO is found. We 
finally emphasize, however, that in order to be able to use our results 
for obtaining truly physical predictions, the NLO corrections to the resolved 
part of the cross section will also have to be calculated in the future.
\section*{Acknowledgements}
The work of one of us (DdF) was partially supported by the World Laboratory.

\end{document}